\documentclass[useAMS,usenatbib]{mn2e}
\usepackage{hyperref}
\usepackage{epsfig}
\usepackage{epstopdf}
\usepackage{graphicx}
\usepackage{amssymb}
\usepackage{amsmath}
\usepackage{graphicx}
\usepackage{subcaption}
\usepackage[utf8]{inputenc}
\usepackage[export]{adjustbox}
\usepackage{wrapfig}
\usepackage{textcomp}
\usepackage[usenames,dvipsnames]{color}
\usepackage{gensymb}
\usepackage{booktabs}
\usepackage{amsmath}
\usepackage{lscape}
\usepackage[english]{babel}
\usepackage{graphicx}
\usepackage{float}
\usepackage{fixltx2e}
\usepackage{soul}
\usepackage[normalem]{ulem}
\usepackage{ulem}

\DeclareMathAlphabet{\mathbbold}{U}{bbold}{m}{n}
\title[Photo-z's and Gaussian Mixed Models]{Augmenting machine learning photometric redshifts with Gaussian mixture models}
\author[Peter Hatfield]{P. W. Hatfield$^{1}$\thanks{peter.hatfield@physics.ox.ac.uk}, I. A. Almosallam$^{2}$, M. J. Jarvis$^{1,3}$, N.Adams$^{1}$, R.A.A. Bowler$^{1}$,\newauthor Z.Gomes$^{1}$, S. J. Roberts$^{4}$, C.Schreiber$^{1}$\\
$^{1}$Astrophysics, University of Oxford, Denys Wilkinson Building, Keble Road, Oxford, OX1 3RH, UK\\
$^{2}$Saudi Information Technology Company, Riyadh 12382, Saudi Arabia\\
$^{3}$Department of Physics, University of the Western Cape, Bellville 7535, South Africa\\
$^{4}$Department of Engineering Science, University of Oxford, Parks Road, Oxford, OX1 3PJ, UK\\
}

\begin{document}


\pagerange{\pageref{firstpage}--\pageref{lastpage}} \pubyear{2020}

\maketitle

\label{firstpage}

\begin{abstract}

Wide-area imaging surveys are one of the key ways of advancing our understanding of cosmology, galaxy formation physics, and the large-scale structure of the Universe in the coming years. These surveys typically require calculating redshifts for huge numbers (hundreds of millions to billions) of galaxies - almost all of which must be derived from photometry rather than spectroscopy. In this paper we investigate how using statistical models to understand the populations that make up the colour-magnitude distribution of galaxies can be combined with machine learning photometric redshift codes to improve redshift estimates. In particular we combine the use of Gaussian Mixture Models with the high performing machine learning photo-z algorithm GPz and show that modelling and accounting for the different colour-magnitude distributions of training and test data separately can give improved redshift estimates, reduce the bias on estimates by up to a half, and speed up the run-time of the algorithm. These methods are illustrated using data from deep optical and near infrared data in two separate deep fields, where training and test data of different colour-magnitude distributions are constructed from the galaxies with known spectroscopic redshifts, derived from several heterogeneous surveys. 

\end{abstract}


\section{Introduction}

Many of the current key open questions in cosmology and extragalactic astronomy require extremely wide (probing large areas of the sky/volumes of the Universe) and deep (probing to extremely faint and/or distant sources) galaxy surveys to answer e.g. observations with Euclid (\citealp{Laureijs2011}) and the Vera C. Rubin Observatory (Rubin, formerly the Large Synoptic Survey Telescope, LSST, LSST Science Collaboration, 2009\nocite{Collaboration2009}). To probe the time evolution/third dimension of the Universe, estimates of the redshift of each galaxy are typically required.

Galaxy (and Active Galactic Nuclei, AGN) redshifts can be calculated from their spectrum in two main ways, from spectroscopy or from photometry. Spectroscopic redshifts (`spec-z's') are calculated by measuring the wavelength of a known spectral (normally emission) line or feature, and comparing it to the known rest frame wavelength of the line or feature. Photometric redshifts (`photo-z's') are calculated by measuring the brightness of the galaxy in $N$ broad wavelength ranges, and mapping these brightnesses onto a redshift. Typically spec-z's are far more precise than photo-z's, but can only be measured for much smaller populations of galaxies as spectroscopic observations are more costly, and generally reach shallower depths, see \citet{Fernandez-Soto2001} for a comparison of the strengths and weaknesses of both classes of measurement.

Photometric redshifts themselves can be calculated in two main ways; `template fitting' methods and `machine learning' methods. Template fitting methods are essentially `theory' based methods - we attempt to use our understanding of the physics behind galaxy spectral energy distributions to map photometry to a spectrum. This in practice can take a number of forms, but typically consists of using a number of model template spectra (either using synthetic spectra or spectra extracted at low-redshift from galaxies with observations at many wavelengths), and using a $\chi^2$-minimisation-like method to find the `best' redshift. Notable template-fitting based codes include Photometric Analysis for Redshift Estimate (LePhare, \citealp{Arnouts1999,Ilbert2006}), Bayesian Photometric Redshifts (BPZ, \citealp{Benitez2000,Benitez2004,Coe2006}), the Zurich Extragalactic Bayesian Redshift Analyzer (ZEBRA, \citealp{Feldmann2006}), EAzY (\citealp{Brammer2008}) and Phosphoros (Paltani et al. in prep). Machine learning photo-z methods are typically entirely data based; a machine learning algorithm is given a set of galaxies with photometry and known (usually spectroscopic) redshifts, and is then tasked with predicting the redshift of galaxies without known spec-z's. Widely used machine learning photo-z codes include Artificial Neural Network Redshifts (ANNz2, \citealp{Collister2004,Sadeh2016}), Trees for Photo-Z (TPZ, \citealp{CarrascoKind2013}), Self Organizing Map Redshifts (SOMz, \citealp{CarrascoKind2013}), Machine-learning Estimation Tool for Accurate PHOtometric Redshifts (METAPHOR, \citealp{Cavuoti2017}), FRANKEN-Z (Speagle et al., in prep.\footnote{https://github.com/joshspeagle/frankenz}) and many more. The resulting photometric redshifts are required for a variety of science goals (see Desprez et al., 2020, for a recent photo-z code comparison in the context of Euclid objectives, and \citealp{Schmidt2020} in the context of Rubin), but one of the most important, with the most stringent requirements, is weak lensing, where the matter power spectrum is measured by the shear on galaxy shapes, but accurate unbiased redshift estimates are needed for unbiased cosmological inferences, e.g. \citet{Banerji2008,Abdalla2008,Hearin2010,Hildebrandt2017,Hoyle2018}. Finally \citet{Salvato2019} present a comprehensive review of contemporary photometric redshift methods, applications and challenges, and a discussion of what advanced approaches must be developed for surveys to best meet their scientific goals in the future.

In this work we consider how using Gaussian Mixture Models (GMMs), a Bayesian approach to dividing a set of objects into sub-populations  can support the machine learning photo-z code GPz (\citealp{Almosallam2016a,Almosallam2016b}) - although the approach could be employed with other algorithms. In particular we investigate using GMMs to a) help account for the different colour space distributions of the training and test data and b) exploit the fact that galaxies and AGN naturally fall into different populations. The approach has some similarities to that described in \citet{Fotopoulou2018}, who divide their galaxies into separate galaxy populations as part of the photometric redshift calculation process (but for a template fitting method), and also \citet{Lima2008}, who use weighting schemes to estimate the redshift distribution, accounting for differences in colour distributions to a reference sample.

The structure of this paper is as follows. In Section 2 we describe the algorithms used in this study, namely GPz and a GMM algorithm, and the data used. In Section 3 we discuss the methods developed. In Section 4 we present our results, discuss the consequences in Section 5, concluding in Section 6.

\section{Preliminaries} \label{sec:Preliminaries}

\subsection{GPz}

GPz is a machine learning regression algorithm originally developed for the problem of calculating photometric redshifts; the details of the algorithm and the key developments in machine learning (ML) theory are described in \citet{Almosallam2016a,Almosallam2016b}. The algorithm is `sparse Gaussian process' (GP) based, e.g. see \citet{Rasmussen2006}. A Gaussian process is a stochastic process with a random variable defined at each point in a space of interest, such that any linear combinations of random variables from different points has a Gaussian distribution; essentially an un-parametrised continuous function defined everywhere with Gaussian uncertainties. GPs are very flexible class of supervised non-linear regression algorithm that make very few explicit parametric assumptions about the nature of the function. For this reason they are well-suited for modelling complex non-linear mappings like photometric redshifts\footnote{Photometric redshift mappings are likely not perfectly represented by a Gaussian Process, but this is likely true of any machine learning algorithms applied to any real-world problem. } (\citealp{Rasmussen2006,Bonfield2010,Almosallam2016a}). A GP based ML algorithm will typically take some set of data over the parameter space of interest and in some sense try and find the Gaussian process that was most likely to have produced the data - and then make predictions for other parts of parameter space based on that. The key features introduced by GPz include a) implementation of a sparse GP framework, allowing the algorithm to run in $O(nm^2)$ instead of $O(n^3)$ (where $n$ is the number of samples in the data and $m$ is the number of basis functions), b) a `cost sensitive learning' framework where the algorithm can be tailored for the precise science goal\footnote{For example, if one is only interested in a science case that requires getting accurate redshifts for $z<1$ galaxies, and such science is insensitive to poor predictions at higher redshifts. In this case the learning cost function would include no penalty for getting $z>1$ galaxy redshifts wrong.}, and c) properly accounting for uncertainty contributions from both variance in the data as well as uncertainty from lack of data in a given part of parameter space (by marginalising over all the GPs that could have produced the data). GPz was further tested and developed in \citet{Gomes2017} who measured the improvement that could be achieved by also including near-infrared bands and angular sizes, as well as introducing a post-processing calibration that reduced the bias (the difference between the true/spectroscopic redshift, and the photo-z estimate, $z_{\textrm{spec}}-z_{\textrm{phot}}$). \citet{Duncan2018} introduced combining GPz with template-based photometric redshifts using a hierarchical Bayesian model that gave better photometric redshifts than ML or template fitting alone could have produced (a hybrid approach was also considered in Desprez et al., 2020). GPz is also beginning to be used in other astronomy and physics applications e.g. building surrogate models for and quantifying the uncertainty on inertial confinement fusion experiments (\citealp{Hatfield2020}) and orbital dynamics (\citealp{Peng2019}).

Two key deficiencies of GPz as applied to photo-z's are a) GPs ordinarily only produce Gaussian uncertainties, whereas the true probability distribution of a galaxy's redshift based on its photometry is typically not Gaussian\footnote{GPs can produce more complex pdfs, but this requires use of a `warped' GP, a mixture of GPs, or a `mixture density' GP }, and b) typically the target galaxy population has a different colour distribution than the colour distribution of the training set, which introduces biases (a problem common to all machine learning based approaches). In this paper we attempt to account for and mitigate against these difficulties.

Unless otherwise stated, we use the settings in \autoref{table-settings} (see \citealp{Almosallam2016a,Almosallam2016b} for precise definitions and interpretations).

 \begin{table*}
\caption{Parameter setting of GPz.}
\begin{center}
\begin{tabular}{| l | l | l |}
    Parameter 	&	Value		&	Description\\	\hline
	$m$			&	500		&	Number of basis functions; complexity of GP, in general higher $m$ is more accurate but longer run time\\
	maxIter		&	500		&	Maximum number of iterations\\
	maxAttempts	&	50		&	Maximum iterations to attempt if there is no progress on the validation set\\
	method		&	GPVC	&	Bespoke covariances on each basis function\\
	normalize	&	True	&	Pre-process the input by subtracting the means and dividing by the standard deviations\\
	joint		&	True	& Jointly learn a prior linear mean-function\\
  \end{tabular}
\end{center}
\label{table-settings}
\end{table*}

\subsection{GMMs}

\textit{Mixture Models} are probabilistic models for modelling data with sub-populations, where the observed data does not identify which population a datum is from. An everyday example of a Mixture Model could be length of publication measured in number of words; the distribution would have separate populations of letters, journal articles and books with very different word length, but typically it would not be possible to separate out short articles from long letters etc. Common astronomical examples include identifying star clusters, identifying populations in surveys, deciding how many classes of a source exist etc. see \citet{Kuhn2017}, and they have also been used for more complex tasks like the identification of strong gravitational lenses (e.g. \citealp{Cheng2020}). Gaussian Mixture Models (GMMs) are a specific type of mixture where each mixture has a Gaussian distribution. GMMs can be viewed as an example of unsupervised machine learning, in that the algorithm is not told in advance what or how many populations there should be or given any examples of members of populations. See also \citet{DIsanto2018} who use a mixture density network to make a GMM of the galaxy redshift posterior.

More formally, we would like to maximize the likelihood distribution:

\begin{equation}
p(\mathbf{X}) = \Pi_{i=1}^{n}p(\mathbf{x}_{i}),
\end{equation}

where $\mathbf{X}=\{\mathbf{x}_{i}\}_{i}^{n}$ is the set of samples in the data set, $\mathbf{x}_{i}\in \mathbb{R}^{d}$ is the $i$-th sample in the data set, $d$ is the dimensionality of the input and $p(\mathbf{x})$ is some multivariate probability density function. Assuming a multivariate normal Gaussian distribution for $p(\mathbf{x})$, the parameters of the distribution that would maximize the likelihood of observing the data, the mean and covariance, can be computed analytically. However, as discussed previously most distributions found in the real world are more complex than a simple unimodal normal distribution. We can model a more complex distribution by assuming that $p(\mathbf{x})$ is the marginal distribution over some latent variable $j$ as follows:

\begin{equation}
p(\mathbf{x}_{i}) = \sum_{j=1}^{k}p(\mathbf{x}_{i}|j)p(j).
\end{equation}

In effect we have modelled the probability density function $p(\mathbf{x})$ as a weightinged sum of Gaussian distributions, i.e. a Gaussian Mixture Model. An important property of GMMs, is that they can, for some number of mixtures $k$, model any non-standard probability distribution. The goal now is to find the $k$ means, covariances and mixture weightings that would maximize the probability of observing the data. This typically cannot be solved analytically, and normally requires optimization techniques. The Expectation Maximization (EM) algorithm is an iterative method that can be used to search for such parameters. However, the EM algorithm is prone to over-fitting especially as the number of mixtures is increased (at the limit when $k=n$, the means will correspond to the data locations and covariances will be close to zero). To overcome this, we use a Variational Bayes (VB, see \citealp{Jordan1999,Jaakkola2000}) approach that puts priors on the means, covariances and mixture weightings to always find the optimal set of mixtures; even if $k$ is set too high it will automatically prune the extra mixtures by setting their mixture weightings to zeros.

\subsection{Data} \label{sec:data}

In a realistic scenario, all galaxies with spectroscopic redshifts would be the training set, and all sources without would be the target test set. Coping with the training and test data sets having different colour-magnitude distributions is a major challenge for ML based photo-z calculations e.g. see \citet{Beck2017}. Unfortunately however, the nature of the problem is that it is difficult/impossible to measure the performance on the galaxies without spec-zs. To overcome this problem, and test how our method performs when the test and training data have different distributions, we construct training and test data sets for which both have spectroscopic redshifts\footnote{This training data is then split 50-50 into what is described as training and validation in \citet{Almosallam2016a,Almosallam2016b} but we shall refer to all the data used in the training process as the training set here.}.

\begin{figure*} 
\includegraphics[scale=0.4]{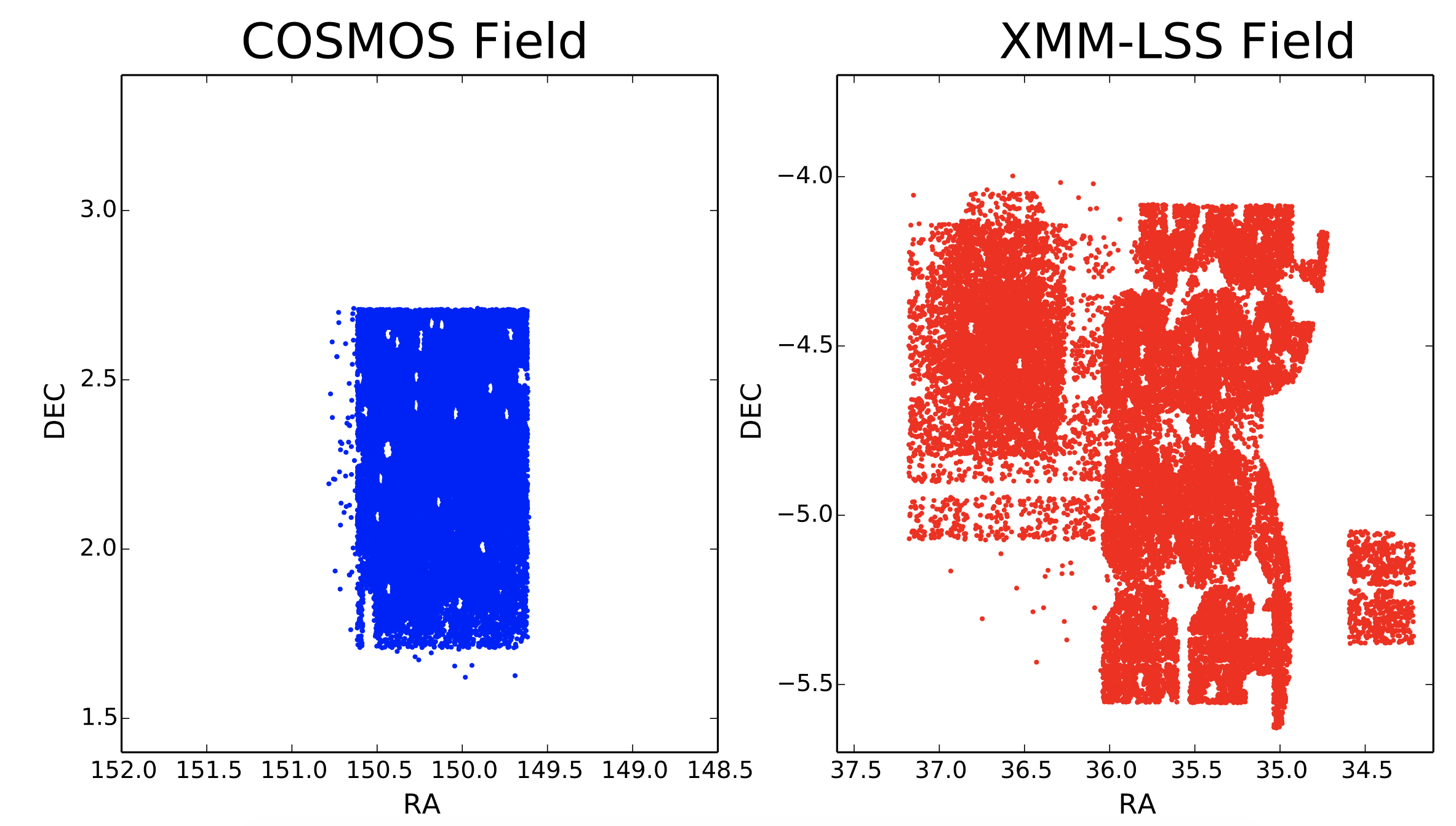}
\caption{Field geometry of galaxies used in this analysis (each of which has a spectroscopic redshift). The `COSMOS' galaxies are used for training, and the `XMM-LSS' galaxies for testing. The unusual field geometries result from the complex ways in which the various photometric and spectroscopic surveys have overlapped and intersected.}
\label{fig:field_geometry}
\end{figure*}

In order to ensure a rigorous test of the methods we consider two separate deep field data sets, but which have similar photometric coverage. Our training data is from the COSMOS field, and the test data from the XMM-Newton Large-Scale Structure (XMM-LSS) field - see figure \ref{fig:field_geometry}. We use the catalogues constructed in \citet{Bowler2020} and \citet{Adams2020}, which, in order to ensure consistency, used identical procedures to extract the photometry across the two fields. Sources were selected in the K$_s$ band, and forced photometry was performed on all the other bands. We use 2'' diameter circular apertures, which had an aperture correction applied by a model generated with PSFEx (\citealp{Bertin2011}) for each band.

For this paper we use the photometry in 10 filters;  $u$ (CLAUDS, CFHT, for both COSMOS and XMM-LSS, \citealp{Sawicki2019}),  GRIZY (HSC-SSP, for both COSMOS and XMM-LSS, \citealp{Aihara2017a}) and YJHK$_{\mathrm{s}}$  (VIDEO-VISTA for XMM-LSS, \citealp{Jarvis2013} and UltraVISTA for COSMOS, \citealp{McCracken2012,Laigle2016}), but to a range of different depths, meaning both the colour space probed and the uncertainty on the photometry are quite inhomogeneous - even though the photometry extraction was done in a very homogeneous manner. The end result is two catalogues that span colour-magnitude space very differently, but with photometry very consistent for comparisons between individual galaxies in the two fields.

Similarly the spectroscopic redshifts come from a range of sources, curated in the same way as in  \citet{Adams2020}\footnote{Which itself was constructed largely similarly to the Catalog of Spectroscopic Redshifts from the Hyper Suprime-Cam Subaru Strategic Program Public Data Release, \url{https://hsc-release.mtk.nao.ac.jp/doc/index.php/dr1_specz/}}. The spec-zs are taken from the VVDS (\citealp{LeFevre2013}), VANDELS (\citealp{McLure2018}; \citealp{Pentericci2018}), Z-COSMOS (\citealp{Lilly2009}), SDSS-DR12 (\citealp{Alam2015}), 3D-HST (\citealp{Skelton2014}; \citealp{Momcheva2016}), Primus (\citealp{Coil2011}; \citealp{Cool2013}), DEIMOS-10K (\citealp{Hasinger2018}) and FMOS (\citealp{Silverman2015}) surveys. We would note that machine learning based photo-z methods are reliant on the accuracy of the spectroscopic redshifts in the training sample. If the spec-zs used in the training process are inaccurate then machine learning methods will simply reproduce the incorrect spec-z values. For this reason we only used the most secure spec-zs that have flags indicating high quality (confidence of $\geq$ 95 per cent). Where a source had a secure spec-z available from more than one survey, the mean of the secure redshifts was used. Furthermore, we found that the Primus spec-zs were often inconsistent with the higher-resolution spec-zs and template-based photo-zs at $z>1$. For this reason we only use the $z<1$ Primus spec-zs. The resulting combination of spectral and photometric data used here is thus similar to that presented in \citet{Adams2020} and \citet{Bowler2020}. 

The XMM-LSS data set constructed includes data from the VANDELS survey, with redshifts in the $z=1-4$ range, which are underrepresented in the COSMOS data for $z>2.5$. This means a) our training and test data have different colour distributions and b) our test data has a high-redshift tail not present in the training data. There are 29,663 galaxies in the COSMOS training set and 24,534 galaxies in the XMM-LSS testing set.

In terms of stellar contamination, as we have tried to ensure that we are only using `secure' redshifts, the vast majority of our sources should be extragalactic. More generally for photo-zs, stars can typically be removed based on a morphological cut (e.g. remove point sources, which will also remove quasars) or with a colour-cut. Conversely, our sample likely has a moderate number of AGN - both in terms of sources that are dominated by AGN light, as well as Seyfert-like galaxies whose photometry has large contributions from both galaxy starlight and a central nucleus. X-ray data is available in both the COSMOS field (\citealp{Marchesi2016}) and the XMM-LSS field (\citealp{Chen2018}), and could in principle be used to identify AGN (considered for some of the sources in this sample in \citealp{Adams2020}). We chose however not to use the x-ray data to remove AGN in this work on the rationale that machine learning methods should in principle be agnostic with regards to whether the source is a galaxy or an AGN - as long as there are similar sources with secure redshifts in the training set it should be possible to give sources accurate machine learning based photo-zs\footnote{There are in practice a few further complications with AGN e.g. temporal variability.}. This conversely is not the case for template based methods; if templates with sufficient AGN contribution are not included in the fitting process then in general Seyfert-like galaxies can receive inaccurate redshifts (a point discussed in greater detail in \citealp{Salvato2019}). The converse strategy of separating a sample into AGN and non-AGN populations in advance is studied in \citet{Norris2019}, who measure photo-z performance when x-ray detected sources are included or not included in their training data. We chose not to study AGN versus non-AGN performance separately in this work in order to focus on how to improve global performance, although making such divisions may be a useful method for some science goals. 

The way the data sets are constructed from a range of photometric and spectroscopic sources means that they do not have single well defined depths, and in general have different colour and redshift distributions. For approximate reference however the 95th percentile faintest training (testing) sources have AB magnitudes $u=26.9$ ($=26.9$), $G=25.6$ ($=25.5$),  $R=24.6$ ($=24.4$),  $I=24.1$ ($=23.9$),  $Z=23.8$ ($=23.6$), $Y_{\textrm{HSC}}=23.6$ ($=23.6$),  $Y_{\textrm{VIDEO}}=23.6$ ($=23.6$), $J=23.3$ ($=23.5$), $H=23.2$ ($=23.4$) and $K_{\mathrm{s}}=23.1$ ($=23.3$). Note these are representative depths of spectroscopic data, not the imaging depths. No cosmology is need to be assumed in this work for any of the photometric or redshift calculations.

\begin{figure} 
\includegraphics[scale=0.45]{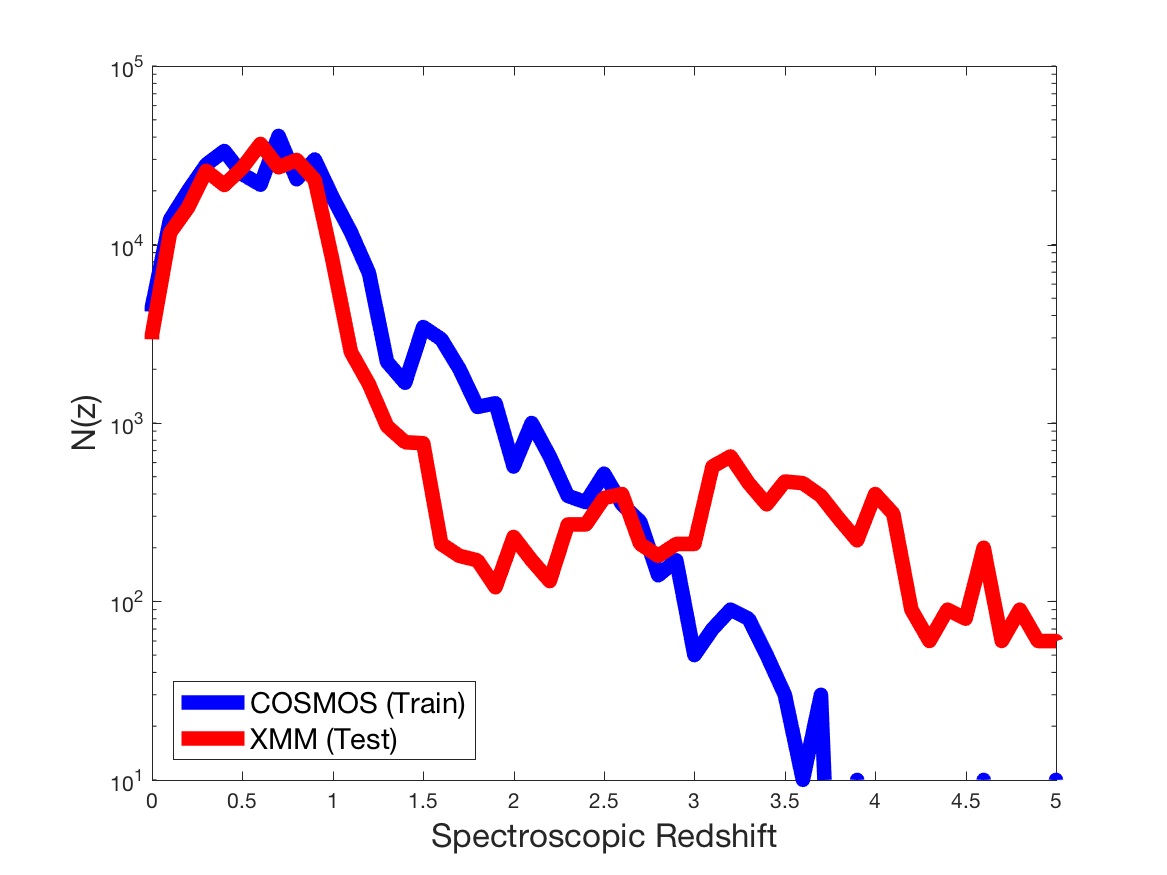}
\caption{The spectroscopic redshift distributions of the galaxies in the two samples used in this study, the `COSMOS'/training data, and the `XMM-LSS'/testing data.}
\label{fig:redshift_distribution}
\end{figure}

\section{Methods} \label{sec:methods}

\subsection{Generalised Cost-Sensitive-Learning} \label{sec:GCSL_description}

A key property of GPz is the ability to tailor the weighting for different science applications. \citet{Almosallam2016a} presented three different weightings, `Normal', `Normalised', and `Balanced'. Normal weighting weightings each galaxy the same (the `null' option), Normalised weightings each galaxy in the training set by $1/(1+z_{\mathrm{spec}})$, and Balanced weightings each galaxy inversely proportional to the number of galaxies with that redshift in the training set (e.g. galaxies with under-represented redshifts are upweightinged).

Here we add to this Generalised Cost-Sensitive-Learning (GCSL) weighting. We model the training set colour-magnitude space\footnote{10D, one magnitude and 9 colours} with a GMM to find $p_{\mathrm{train}}(\textbf{x})$, and do the same for the colour-magnitude space of the test set of data to find $p_{\mathrm{test}}(\textbf{x})$(where $\textbf{x}$ is the colour-magnitude photometry space of the data). This is essentially using the GMM to model the probability distribution in an unbiased way. We then set the weighting for GPz as:

\begin{equation} \label{eq:weighting}
w_{i}=\frac{p_{\mathrm{test}}(\textbf{x}_i)}{p_{\mathrm{train}}(\textbf{x}_i)} .
\end{equation}

\begin{figure} 
\includegraphics[scale=0.5]{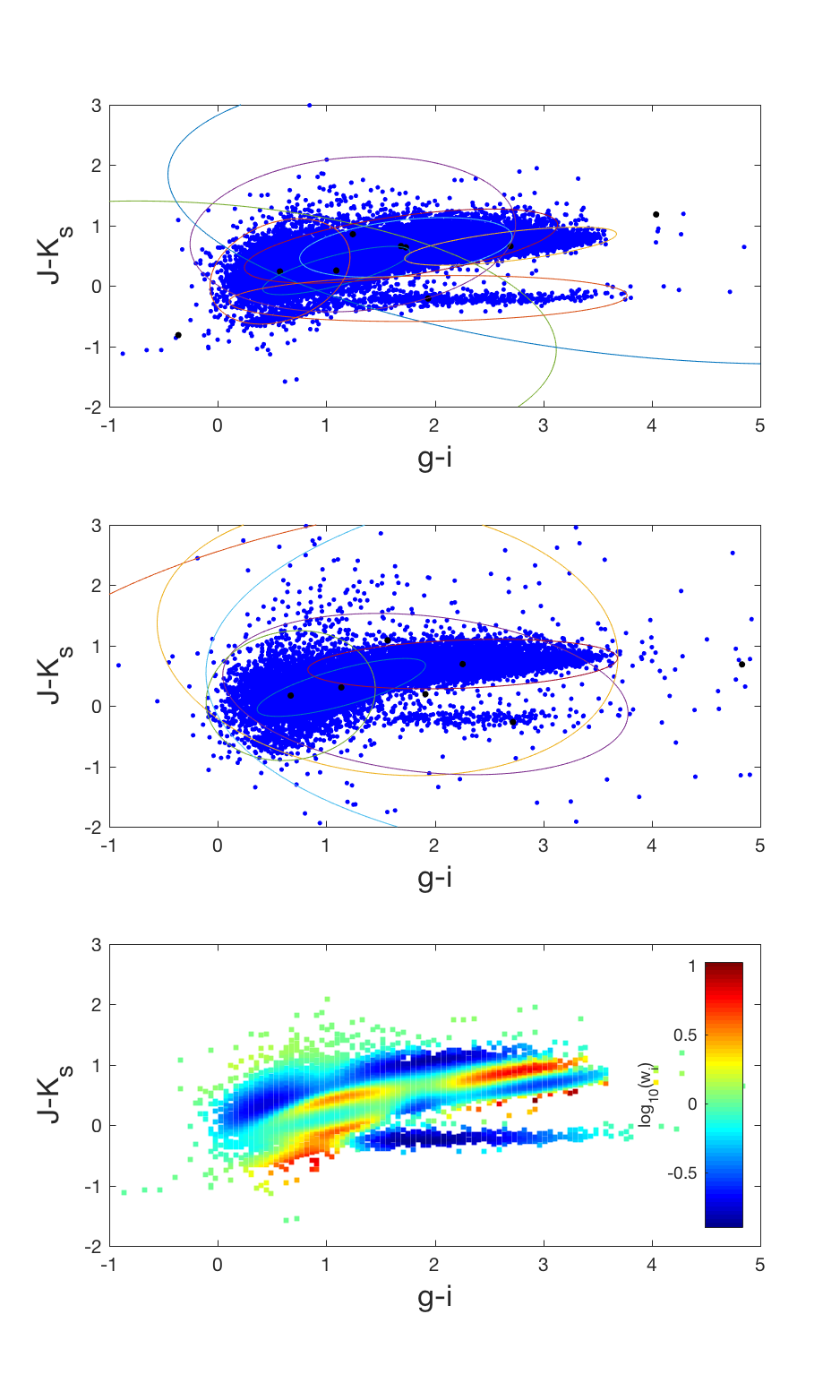}
\caption{The (g-i)-(J-K$_{\mathrm{s}}$) plane for the training (top sub-plot) and test (middle sub-plot) data. Ellipses show the identified mixtures (projected into 2D), with black marks showing the mixture centres. The lower plot shows the $w_{i}$ values for the training data (essentially the ratio of the data density of the top and middle plots). A stellar locus is visible (c.f. Figure 6a in \citealp{Baldry2010}).}
\label{fig:weighting_colour_space}
\end{figure}

\begin{figure} 
\includegraphics[scale=0.4]{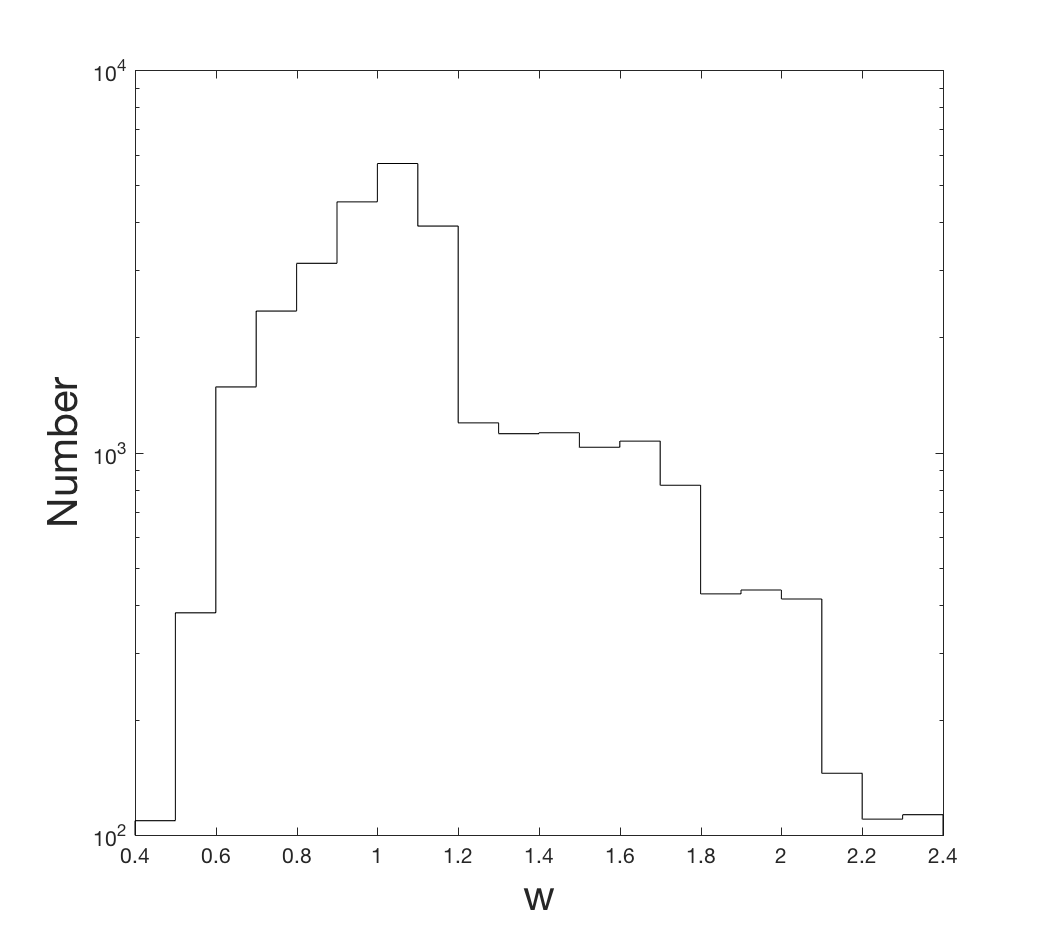}
\caption{Distribution of $w_{i}$ values found (note that $w_{i}$ is capped at 20).}
\label{fig:weighting_distribution}
\end{figure}

This potentially improves over balanced weighting a) because it weightings galaxies in colour space rather than redshift space (different parts of colour space correspond to the same redshift) and b) it accounts for the colour-magnitude distribution of both the training set \textit{and} the test set, rather than just the training set (it essentially generalises the approach used in \citealp{Lima2008} and \citealp{Duncan2018}). A small number of extreme outlier galaxies end up with extremely high weightings which were found to distort the entire training process, so we a) cap the maximum weighting at 20, and b) instead of equation \ref{eq:weighting}, we in practice use $w_{i}=\frac{p_{\mathrm{test}}(\textbf{x}_i)+\epsilon}{p_{\mathrm{train}}(\textbf{x}_i)+\epsilon}$ with $\epsilon=0.01$ to avoid extreme ratios at parts of parameter space with low data densities.  Figure \ref{fig:weighting_colour_space} shows (a 2D projection of) the different colour spaces of the training and test data sets, the mixtures that were identified, and the corresponding $w_{i}$ values. Figure \ref{fig:weighting_distribution} shows a histogram of resulting $w_{i}$ values.

If the test set and the training set have the same colour distribution GCSL reduces to `Normal' weighting as $p_{\mathrm{train}}$ and $p_{\mathrm{test}}$ will be identical and the $w_{i}$ will be equal to one. In this scenario, for rare parts of colour-magnitude space, the performance will be low, but that is unimportant because a proportionately small number of galaxies in the test set will have that colour. Balanced is near-equivalent\footnote{Only near-equivalent, as in \citet{Almosallam2016a} the weighting is based on redshift rather than colour.} to when the test set has a uniform distribution in colour space, or equivalently desiring a homogeneous performance in colour space. `Normalised' makes the science judgement that we value percentage error on $1+z$ rather than absolute error on $1+z$, as opposed to weighting on sample distributions. One could in principle weighting by `Normalised-GCSL' with $w_{i}^{\mathrm{Normalised-GCSL}}=w_{i}^{\mathrm{GCSL}}/(1+z_i)$, although we found in practice this made relatively little difference.

From figure \ref{fig:weighting_colour_space} it can be seen that there is some stellar contamination in the sample (the population at $\mathrm{J}-\mathrm{K}_{\mathrm{s}}\approx-0.2$, see Figure 6a in \citealp{Baldry2010}), despite them nominally being spectroscopically confirmed galaxies (as is likely to be the case at least to some degree in any putative Euclid and Rubin data set). However it can be seen that the GMMs do identify the stars as being a separate component in colour-colour space (and down-weighting them, as there do not appear to be as many in the test spec-z sample), so future work could include a probability of each mixture to correspond to a stellar component.

\subsection{Weighted Validation} \label{sec:validation}

Machine learning algorithms generally divide the labelled data into training and validation data sets when training the algorithm. The algorithm typically fits parameters to the training data, measures how well it does on the validation data, and then updates the complexity of the model (if over/under fitting) appropriately based on this. Since we are ultimately trying to predict the test data, making the validation data look more like the test data might help the model better tune itself. To do this, we take each galaxy with a spectroscopic redshift in the COSMOS data set and probabilistically select it to be either training or validation according to:

\begin{equation}
P_{\mathrm{train}}=\frac{1}{1+w_{i}}
\end{equation}
and
\begin{equation}
P_{\mathrm{valid}}=\frac{w_{i}}{1+w_{i}},
\end{equation}

\noindent where $P_{\mathrm{train}}$ is the probability of the galaxy being assigned to the training data set, $P_{\mathrm{valid}}$ is the probability of the galaxy being assigned to the validation data set and $w_{i}$ is the weighting from equation \ref{eq:weighting}. Weighted Validation is similar in some regards to Generalised Cost-Sensitive-Learning but may potentially cope slightly better with parts of parameter space with low data density.

\subsection{Resampling} \label{sec:resampling}

As already mentioned, GP based ML methods, including GPz, by definition only give Gaussian uncertainties. This is typically not realistic for photometric redshifts. To generate non-Gaussian ML posteriors with GPz we trial a resampling method, where GPz is run a large number of times on slightly perturbed copies of the data to produce a large number of Gaussians, which are then summed to produce a non-Gaussian pdf. This is similar to the photometry perturbations of METAPHOR (\citealp{Cavuoti2017}), and the Monte Carlo approach of FRANKEN-Z (Speagle et al., in prep.).

The resampling method consists of the following steps:
\begin{enumerate}
  \item For each galaxy in the training and test sets, based on the uncertainty on each magnitude, resample a new magnitude value for each band
  \item Train GPz on this resampled set of training data
  \item Produce Gaussian pdfs for the resampled test data
  \item Repeat Steps 1-3 $s$ times
  \item Average the $s$ pdfs produced to obtain the final pdf
\end{enumerate}

This procedure lets us probe all the variation in the data. The hope is that for galaxies near `cliffs', that are assigned one redshift with a very small uncertainty, but are very close in colour-magnitude space to galaxies that assigned very different redshifts also with small uncertainty, that the `cliff' is repositioned slightly each time and the galaxies get more realistic pdfs. This approach can be thought of as a numerical method for GPz with noisy input. GPz conventionally accounts for noisy input, but still produces a single Gaussian. This approach approximates, at the limit of large `$s$', a multi-modal Gaussian for a `noisy' input of variance equal to the perturbation variance. Of the methods suggested in this text, this is by far the most computationally expensive, as it increases the runtime by a factor of $s$. However, because of the sparse framework used, GPz runs very fast and it remains practical to use $s\sim100$ on a laptop (as used in this analysis), and would be viable to use a much higher $s$ on a cluster as each `run' of GPz can be parallelised.

\subsection{Exploiting the Population Structure} \label{sec:divide_description}

Galaxies naturally fall into different populations. We can use the GMM to find natural galaxy populations, and assign probabilities of being in each population. Given a GMM, and for each galaxy a discrete probability density function for being in each of the populations, there are two natural ways to couple this to a ML algorithm, which we discuss below.

In the basic application of GPz, training the algorithm is $O(nm^2)$ in number of samples $n$ and in number of basis functions $m$ used to represent the mapping from photometry to redshift (in general a higher $m$ will give better results, but take longer to train). Here we propose using the GMM to split colour space into $k$ regions, and then running  GPz with $m/k$ basis functions in each region (we call this `GMM-Divide'). This typically will reduce the run time by a factor of $\sim k$ - each region will run $\sim k^2$ times faster, but $k$ regions must be run. Regions are defined by which Mixture has the highest probability for that point in colour-magnitude space (a galaxy is in the region of the population it is most likely to be in). It is technically possible for the region for some Mixtures to be the empty set e.g. tiny Gaussian within a broader one with a much larger amplitude. Regions are thus completely algorithmically determined by the GMM, with no human intervention; see figure \ref{fig:weighting_colour_space}.  We define regions of the colour-magnitude distribution of the test set, and separate both the training and test set based on that region division\footnote{Similar results are achieved when defining the regions based on the training set.}. There are many slight variations on exactly how the GMM algorithm can find populations; we found that letting it select in a magnitude and colour space (see figure \ref{fig:weighting_colour_space}) gave the best results although precise implementation does not appear to make a large difference. The approach of splitting up parameter space into multiple regions has some similarities to the method of \citet{Masters2015}, who use Self-Organising Maps (SOMs) rather than Gaussian Mixture Models as the unsupervised learning model used to divide up colour-magnitude space. There is also some overlap to the approach of separating AGN and non-AGN in advance described in \citet{Norris2019}; if x-ray data is available for the training process, then separating based on whether or not a source is x-ray detected is essentially dividing the sample into two populations based on a flux cut. The GMM here attempts to make such divisions in an unsupervised manner.

The other natural approach, which we investigated, would be to let galaxies be in multiple populations probabilistically e.g. for each population, we train GPz on the entire training set with weightings proportional to how likely they were to have been drawn from that population i.e. $w_{i} = P(\textrm{galaxy $i$ in population $h$})  $. We then calculate a redshift pdf for each galaxy in the training set for each population. These are then summed in proportion to the probability that each galaxy in the test set was in each population (e.g. if there was a 75 percent probability the galaxy was drawn from population 1, and a 25 percent chance it was drawn from population 2, the Gaussians from GPz run for population 1 and population 2 are summed with a 3:1 weighting). However we found that the vast majority of galaxies were in a given population with probability near 1, and that this approach did not typically give better predictions.

\subsection{Predicting $\mathrm{log} (z)$} \label{sec:log}

With the exception of Andromeda and other bodies in the Local Group, all galaxies have a positive redshift. Machine learning based redshift predictions can sometimes have non-trivial fractions of the resulting pdf be negative. One way of coping with this is to effectively set a strong prior that redshift has to be non-negative (essentially cut off the negative part of the pdf). This unfortunately has the result of introducing a bias in the predictions for low redshift galaxies (because this approach means you can only over-estimate, never underestimate). We test the alternative of trying to predict $\mathrm{log} (z)$, which now takes all values, rather than $z$ (considered in Section 3.3 of \citealp{AlmosallamThesis2017}). Note that if $\Theta=\mathrm{log} (z)$, $\mathbbold{E}(z) \neq \exp( \mathbbold{E}(\Theta))$; \citet{AlmosallamThesis2017} show $\mathbbold{E}(z) = \exp( \mathbbold{E}(\Theta)+\frac{1}{2}\mathbbold{V(\Theta)}))$ and $\mathbbold{V}(z) = (\exp( \mathbbold{V}(\Theta))-1) \times (\mathbbold{E}(z))^2$.  

This problem is to some degree similar to the issue of treatment of negative fluxes; fluctuations in the noise can lead to galaxies being assigned negative fluxes through the data reduction process (although hopefully with uncertainties that make the measurements consistent with zero flux). When we train our photometric redshift model, we use log-fluxes i.e. implicitly assuming uncertainties on fluxes are Gaussian in log-space. For negative flux values we resample from a Gaussian centred on the negative value, with standard deviation the associated error, until a positive value is found. This obviously gives a slight bias for the faintest galaxies, however, for these sources the uncertainty on the flux is high and as the uncertainty is used within GPz it has negligible effect. An alternative way to deal with negative fluxes is to use luptitudes (\citealp{Lupton1999}), as used with GPz in Desprez et al., (2020). Fluxes typically have Gaussian uncertainty in linear-space near the detection threshold, but Gaussian uncertainty in log-space for brighter sources. The luptitude transformation essentially transforms the flux in such a way as to smoothly transition between these two regimes. Unfortunately however this requires consistency of detection threshold (which is used in the transformation), and the detection thresholds are not uniform between the different surveys we use to make up our two samples, so a luptitude in one would not necessarily correspond to the same luptitude in another. Both the method employed here and luptitudes are not completely accurate, however the effect on our results is minimal.

\section{Results} \label{sec:corr_functions}

In this section we trial the methods discussed in Section \ref{sec:methods} on the data described in Section \ref{sec:data}.  We calculate photometric redshifts using the following methods:

\begin{itemize}
\item Base performance, `Normal' weighting
\item `Generalised CSL' weighting (section \ref{sec:GCSL_description})
\item Weighted Validation (section \ref{sec:validation})
\item Resampled base performance as per section \ref{sec:resampling} with \indent \indent $s=100$
\item GMM-Divide (section \ref{sec:divide_description})
\item Modelling  $\mathrm{log} (z)$ (section \ref{sec:log})
\item `All' - Weighted Validation',  GMM-Divide and Resampling done simultaneously \footnote{There are a large number of ways that all different approaches could be combined; this method we found was the most logical and highest performing.}
\end{itemize}

\begin{figure*} 
\includegraphics[scale=0.6]{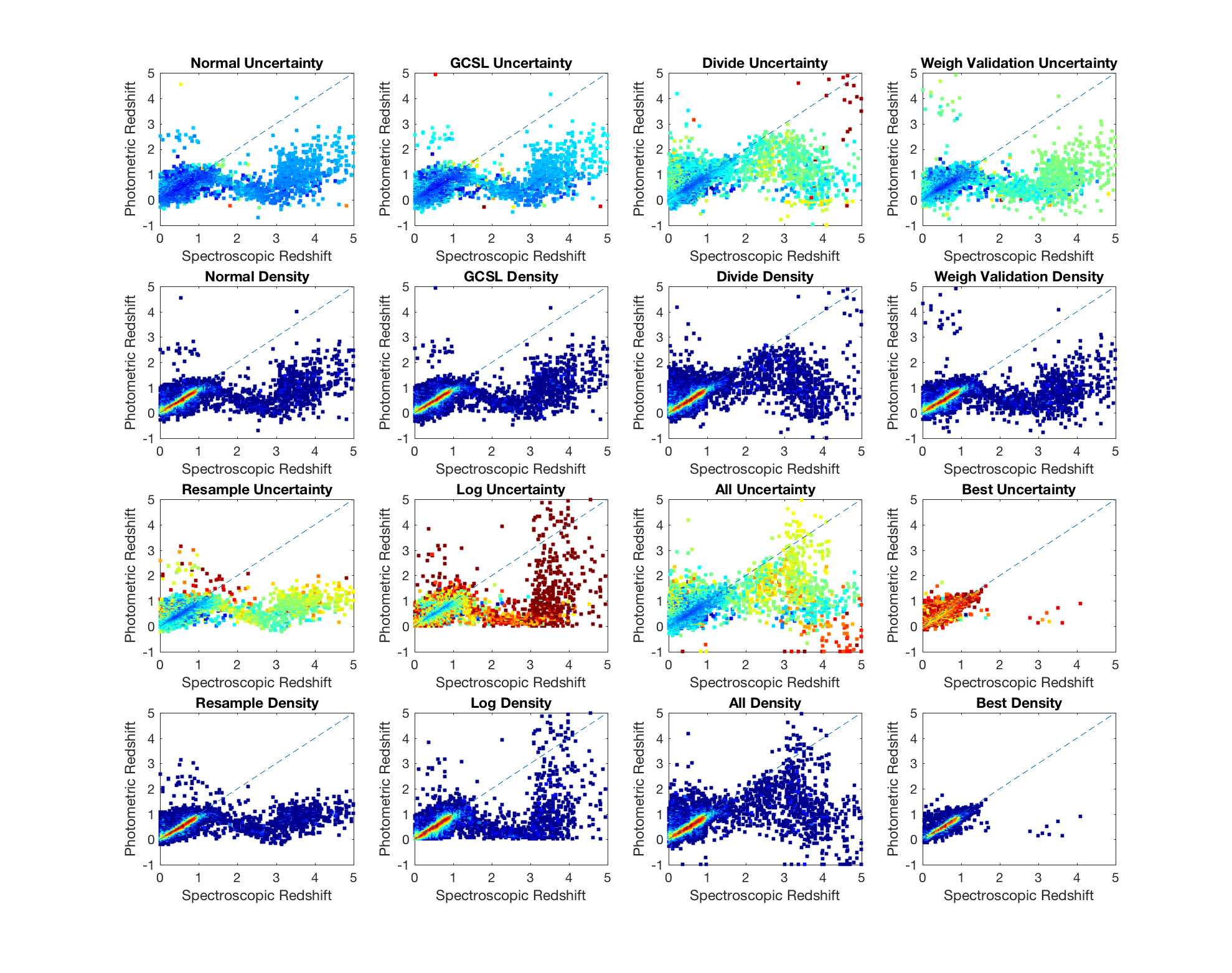}
\caption{Spectroscopic redshift versus photometric redshift for the Normal implementation, Generalised-CSL, GMM-Divide and Weighing Validation methods (top two rows, going left to right) and the resampling, log, `All' and `Best' methods (bottom two rows, going left to right). The first and third rows are coloured by the predicted uncertainty, the second and fourth by the data density.}
\label{fig:z_z}
\end{figure*}

Figure \ref{fig:z_z} shows spec-z versus photo-z for these methods, with varying performance. We also show for reference results if only the 70 percent of data with lowest predicted uncertainty from `All' is used (`Best').

\subsection{Metrics} \label{sec:metrics}

\begin{figure} 
\includegraphics[scale=0.75]{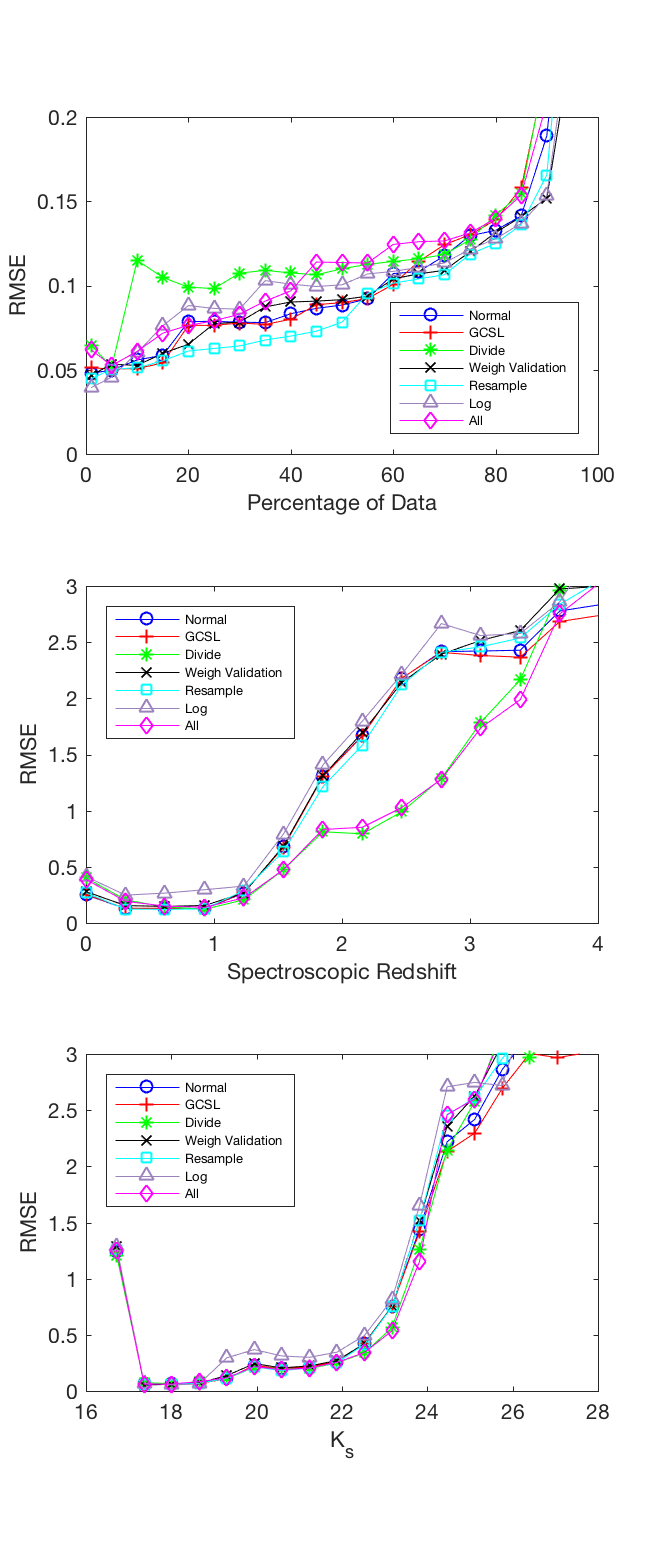}
\caption{Root mean squared error on on the photometric redshifts, as a function of a) percentile of the data (top plot, zero with the smallest uncertainties, 100 the largest), b) spectroscopic redshift (centre plot) and c) $K_{\mathrm{s}}$ band magnitude (bottom plot).}
\label{fig:RMSE}
\end{figure}

\begin{figure} 
\includegraphics[scale=0.5]{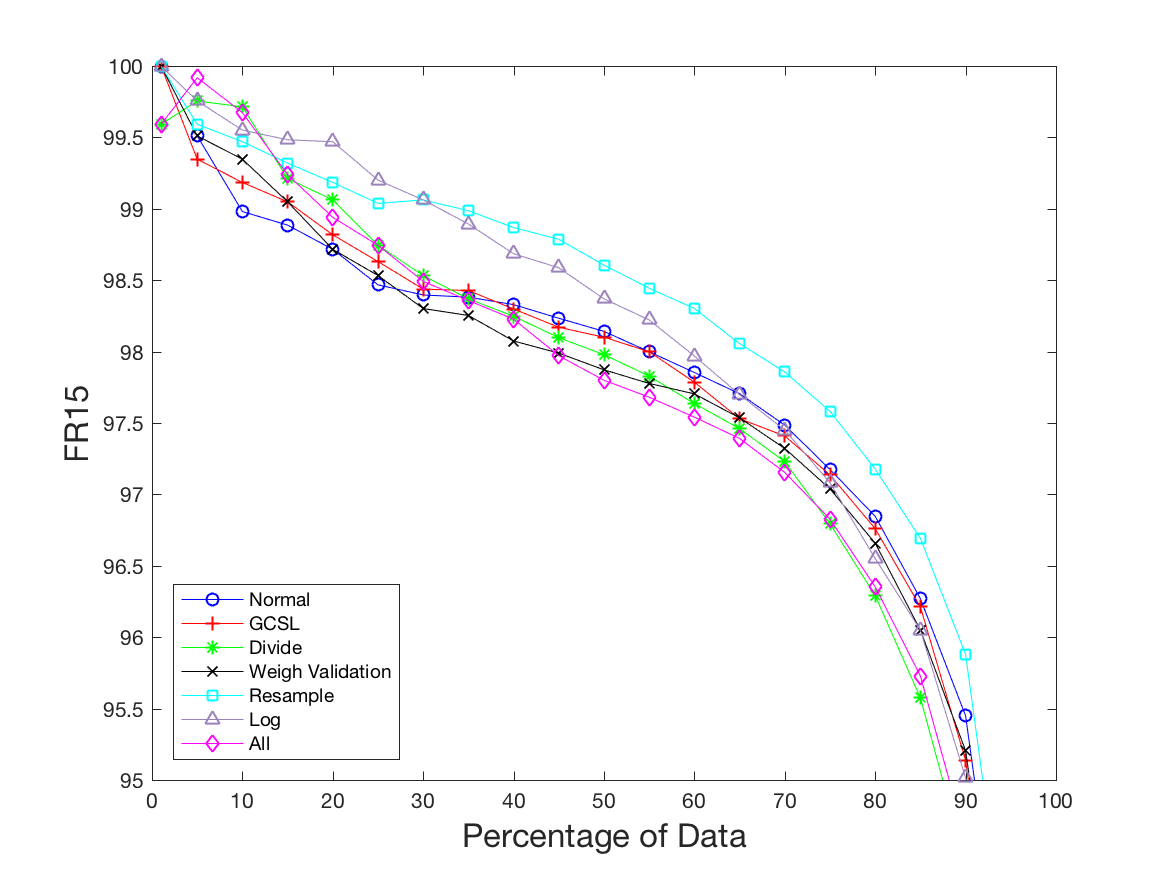}
\caption{FR15 on on the photometric redshifts, as a function of percentile of the data (zero with the smallest uncertainties, 100 the largest).}
\label{fig:FR15}
\end{figure}

\begin{figure} 
\includegraphics[scale=0.5]{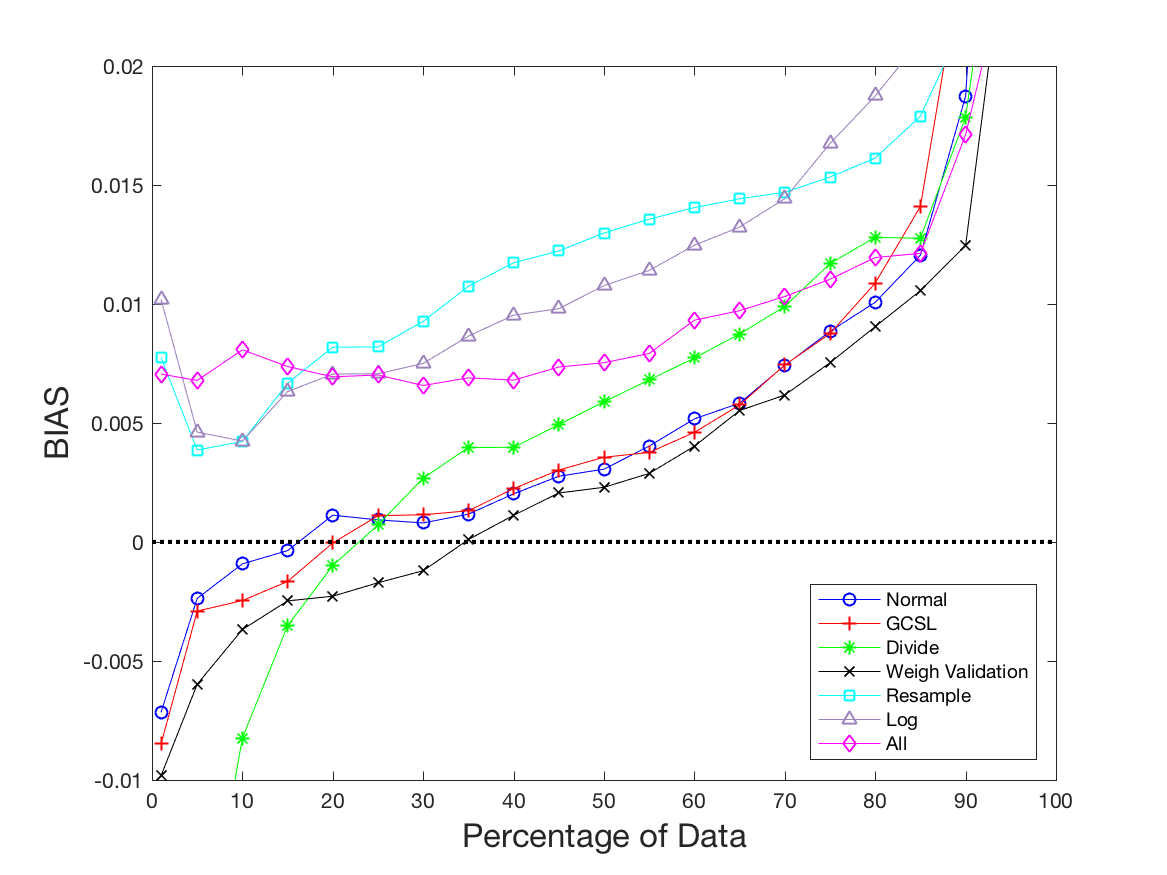}
\caption{Bias on on the photometric redshifts, as a function of percentile of the data (zero with the smallest uncertainties, 100 the largest).}
\label{fig:bias}
\end{figure}

Figures \ref{fig:RMSE}, \ref{fig:FR15}, and \ref{fig:bias} compare the performance of our methods\footnote{Comparing the mode of the photo-z pdf to the true spectroscopic redshift where appropriate} as measured by the root mean squared error (RMSE), bias ($z_{\textrm{spec}}-z_{\textrm{phot}}$), and fraction of sources within 15 percent of the true value (FR15), see table 1 in \citet{Gomes2017}. These quantities are expressed as a function of `fraction of the data' (the data is divided into bins of `error bar size', so ten percent corresponds to a bin of the best tenth of galaxies in terms of uncertainty size etc.). For RMSE, we also show the data as a function of spectroscopic redshift. 

RMSE is approximately 0.05 for the data with the smallest uncertainties, and increases to about 0.25 for the data with the largest uncertainties. As a function of (spectroscopic) redshift we find the RMSE $\approx0.2$ at $z\sim0.75$, and $\approx3$ at high redshift. As a function of $K_{\mathrm{s}}$-band magnitude the RMSE is between 0 and 0.5 for most magnitudes, but rises rapidly for $K_{\mathrm{s}}>23$ and $K_{\mathrm{s}}<17$ galaxies, where there is less training data\footnote{We considered RMSE as a function of $K_{\mathrm{s}}$ because that was the band that detections were made in.}.  FR15 varies from essentially 100 percent for the data with the smallest uncertainties, to around 85 percent for the data with the largest uncertainties, dropping off sharply for the final 30 percent of the data, as one would expect. Bias is less than 0.02 for most of the data, apart from the 20 percent with the largest uncertainties. The different methods gave moderate variability; in particular `Resampling' improved RMSE and FR15, but increased the bias. `Weighted validation' was the only method to largely improve the bias.

\begin{figure*} 
\includegraphics[scale=0.5]{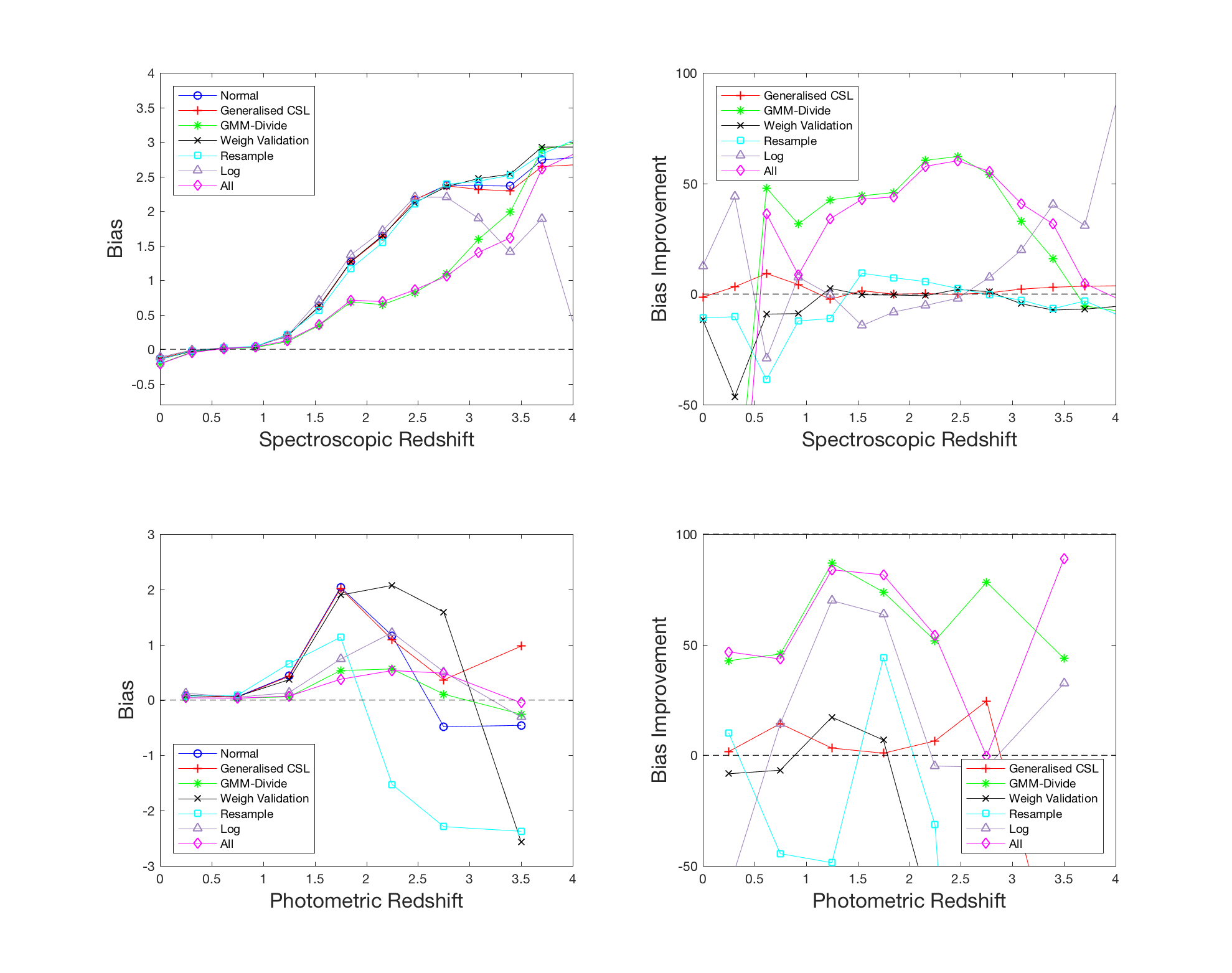}
\caption{Bias on photometric redshift estimation (left column), and improvement in bias (right column) compared with `Normal'; 0 percent is no change in bias and 100 percent is complete removal of bias.  Top row shows the results as a function of spectroscopic redshift, bottom row shows as a function of predicted photometric redshift.}
\label{fig:bias_four_improvement}
\end{figure*}

Figure \ref{fig:bias_four_improvement} shows a) the bias on the photo-z's as a function of both spec-z, and photo-z, and b) the relative improvement in bias, compared with `Normal' weighting. It can be seen that the bias is between -0.2 and +0.2 for $0<z<1$, but steadily increases for higher redshifts). The redshifts are essentially biased towards where most of the data is, which gives rise to the redshift dependence (it is also essentially impossible to underestimate the redshifts of very low redshift sources because $z>0$). Whether bias as a function of spec-z or photo-z is more important depends on science goal. For a real science problem, only the photo-z's of the test data will be available, but bias as a function of spec-z can also be relevant depending on whether false-positives or false-negatives are more relevant for a high-redshift science goal etc.

It can be seen in figure \ref{fig:bias_four_improvement} that `Divide' reduces the bias at higher redshifts (which is inherited by `All'). `Log' slightly improved the bias at the highest redshifts, and `Resample' impaired the bias (as a function of photo-z), but largely the methods apart from `Divide' and `All' didn't give any major improvements.

\subsection{Quality of Probability Distributions} \label{sec:pdfs}

\begin{figure} 
\includegraphics[scale=0.5]{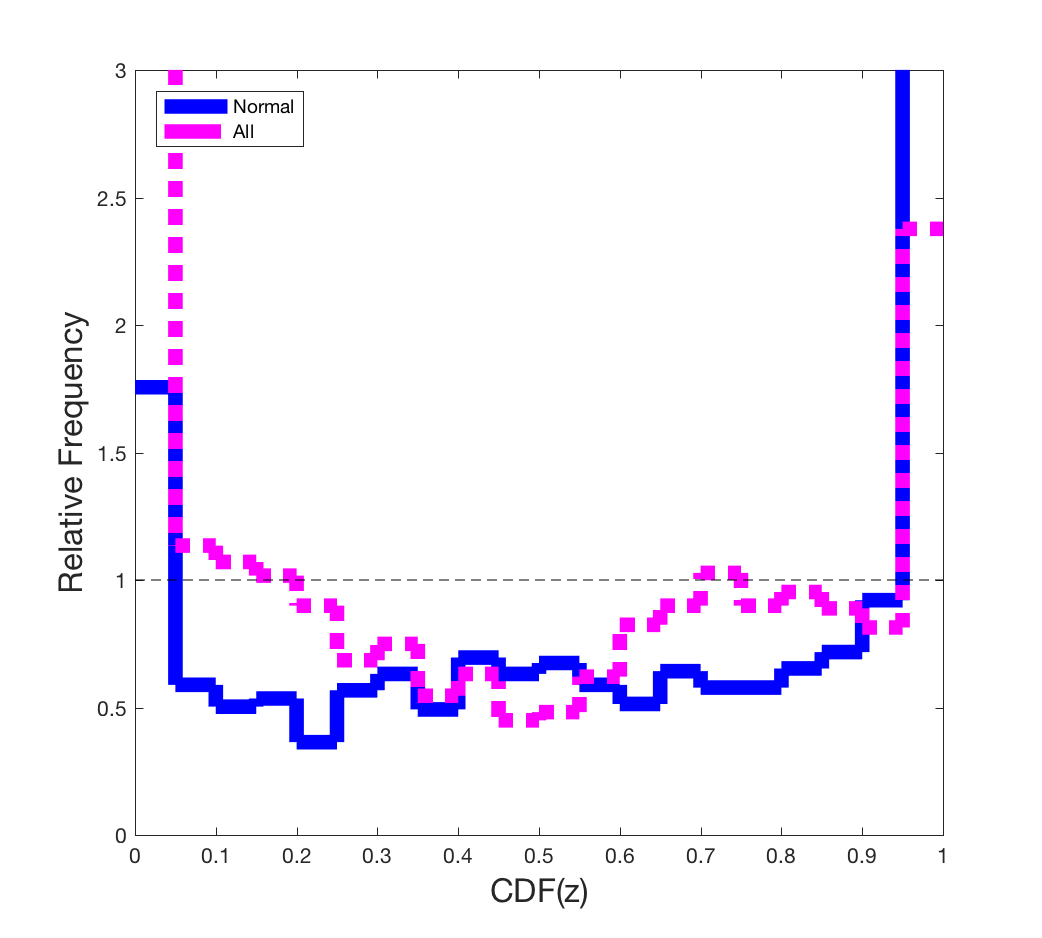}
\caption{PIT plot for all the test data for `Normal', and `All' methods for photometric redshifts $z>1$. Probability distributions are perfectly calibrated if they lie on the horizontal dashed line.}
\label{fig:PIT}
\end{figure}

Probability Integral Transform (PIT) plots can be used to compare how well calibrated these pdfs are (\citealp{DIsanto2018}). The plot essentially shows a histogram of cumulative distribution function values at the spectroscopic redshift (e.g. for each galaxy calculate what fraction of the pdf is less than the true value, and plot a histogram of these values). Figure \ref{fig:PIT} shows the PIT plot for all $z>1$ test data, for both `Normal' and `All'. The asymmetric `U' shapes show that the pdfs are slightly over-narrow and biased - but the fact that the `All' curve is closer to a flat line (which would correspond an unbiased pdf) shows the quality of the pdfs has indeed been improved at high redshift. For $z<1$ the PIT plots for `Normal' and `All' are essentially identical.

\subsection{Population Redshift Distribution} \label{sec:population_distribution}

\begin{figure} 
\includegraphics[scale=0.5]{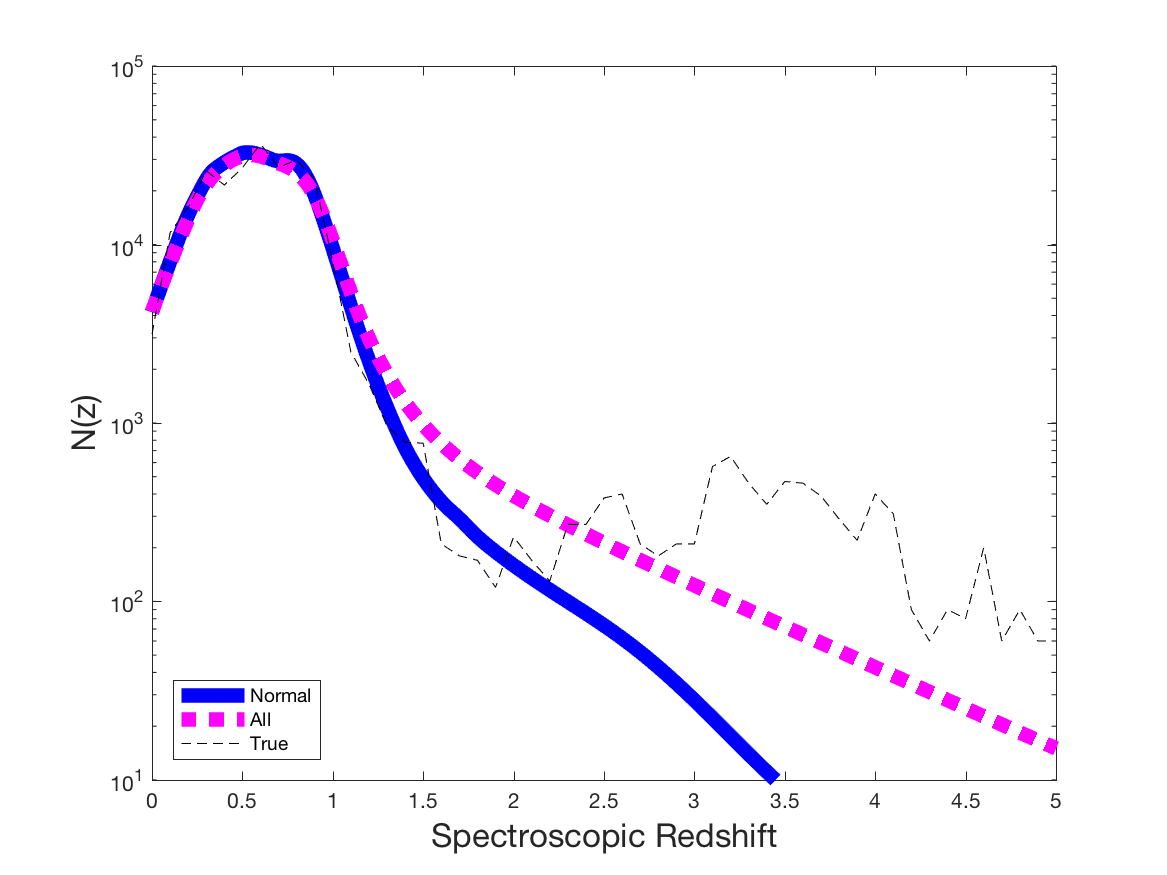}
\caption{The true underlying redshift distribution (black dashed line), and the stacked pdfs from the a) 'Normal' (full blue line) and b) 'All' (dotted magenta line) methods. It can be seen that  'All' better captures the high redshift distribution. }
\label{fig:stacked}
\end{figure}

Finally we plot the summed probability distributions\footnote{More sophisticated methods for estimating the sample redshift distribution do exist, but we do not consider here, e.g. \cite{Leistedt2016}.} for both `Normal' and `All' to reconstruct estimates of redshift distributions for the test data in figure \ref{fig:stacked}. Both `Normal' and `All' get the lower redshift `hump' correct, but both struggle to identify the higher redshift hump. This is not surprising, given the training data, but it does show that using `All' does manage to $\sim$triple the estimated number of high redshift galaxies, even if this estimate is still below the true number.

\section{Discussion}

Our results show that it is possible to obtain significant improvements to the performance of GPz with comparatively little input (and no extra data), taking into account the differences in colour-magnitude distribution of the training and test data. Although we have discussed in the context of GPz, these methods could be easily extended to any other ML photo-z code. In particular GMM-Divide is easily implemented and gives large improvements, and may be particularly valuable for Euclid and Rubin, where there will be billions of sources, and colour-magnitude space could be profitably split up into hundreds of sections, each still containing $\sim 10^5 - 10^7$ galaxies. The methods described here could also be combined with the pdf post-processing method described in \citet{Gomes2017}.

Improvements at higher redshifts seem to largely come from `Divide' (modelling different parts of colour-magnitude space separately based on a GMM). This is likely because the `Divide' method identifies a population that corresponds to higher-redshift galaxies. Because each population gets the same share of basis functions, this population gets more basis functions than it normally would in the straightforward implementation of GPz, and can be modelled more accurately. The lower-redshift sources get fewer basis functions than they otherwise would, but still get high performance, as they were only getting a large number of basis functions as it was where the bulk of the data was. An interesting comparison for future analyses where x-ray and/or radio data was available might be to compare the merits and demerits of unsupervised divisions like that discussed in this work, versus cuts explicitly designed to separate AGN and non-AGN.

Resample most improved the RMSE and FR15, but adversely affects the bias. Weight Validation was the best at improving bias. Using `Divide', `Resample' and `Weight Validation' together (`All') thus seemed the best way to improve both bias, scatter, and behaviour across all redshifts, and this is borne out by the `All' results for figure \ref{fig:bias_four_improvement}.

Modelling  $\mathrm{log} (z)$ successfully avoids negative redshifts, but typically led to poorer results. It also to some degree pushes the issues at $z=0$ to $z=\infty$; galaxies up-scattered to high $\mathrm{log} (z)$ end up with predicted redshifts in the hundreds. If choosing to keep with modelling $z$ rather than $\mathrm{log} (z)$ it, as discussed, can be tempting to simply `cut off' the negative probability (e.g. declare the final pdf for the galaxy's redshift to be a Gaussian multiplied by a step-function). This can be an acceptable solution depending on science goals, but as discussed it does have the unfortunate feature of biasing results near $z=0$ e.g. making that cut forces the redshifts to be over estimated. This can be overcome if one requires the whole redshift distribution; some galaxies being assigned negative redshifts can be accepted, and then accounted for when finding the redshift probability distribution for the whole population e.g. with a hierarchical Bayesian model as per FRANKEN-Z  (Speagle et al., in prep.).

Figure \ref{fig:PIT} suggests that `All' improves the calibration of the redshift pdfs for $z>1$.  As noted however the asymmetric `U' shapes show that the pdfs are generally slightly too narrow.  Having accurate pdfs (as opposed to simply point estimates) is essential for many applications, including luminosity functions e.g. \citet{Lopez-Sanjuan2017}. More generally at high redshift it is sometimes found that template-fitting methods can under-estimate uncertainty (\citealp{Dahlen2013,Salvato2019})\footnote{Although many works now mitigate against this e.g. \citet{Buchner2015}.}, with machine learning methods usually producing more realistic pdfs (e.g. \citealp{Brescia2019}).

\subsection{Comparison to Template Based Methods} \label{sec:template_methods}

\begin{figure*} 
\includegraphics[scale=0.6]{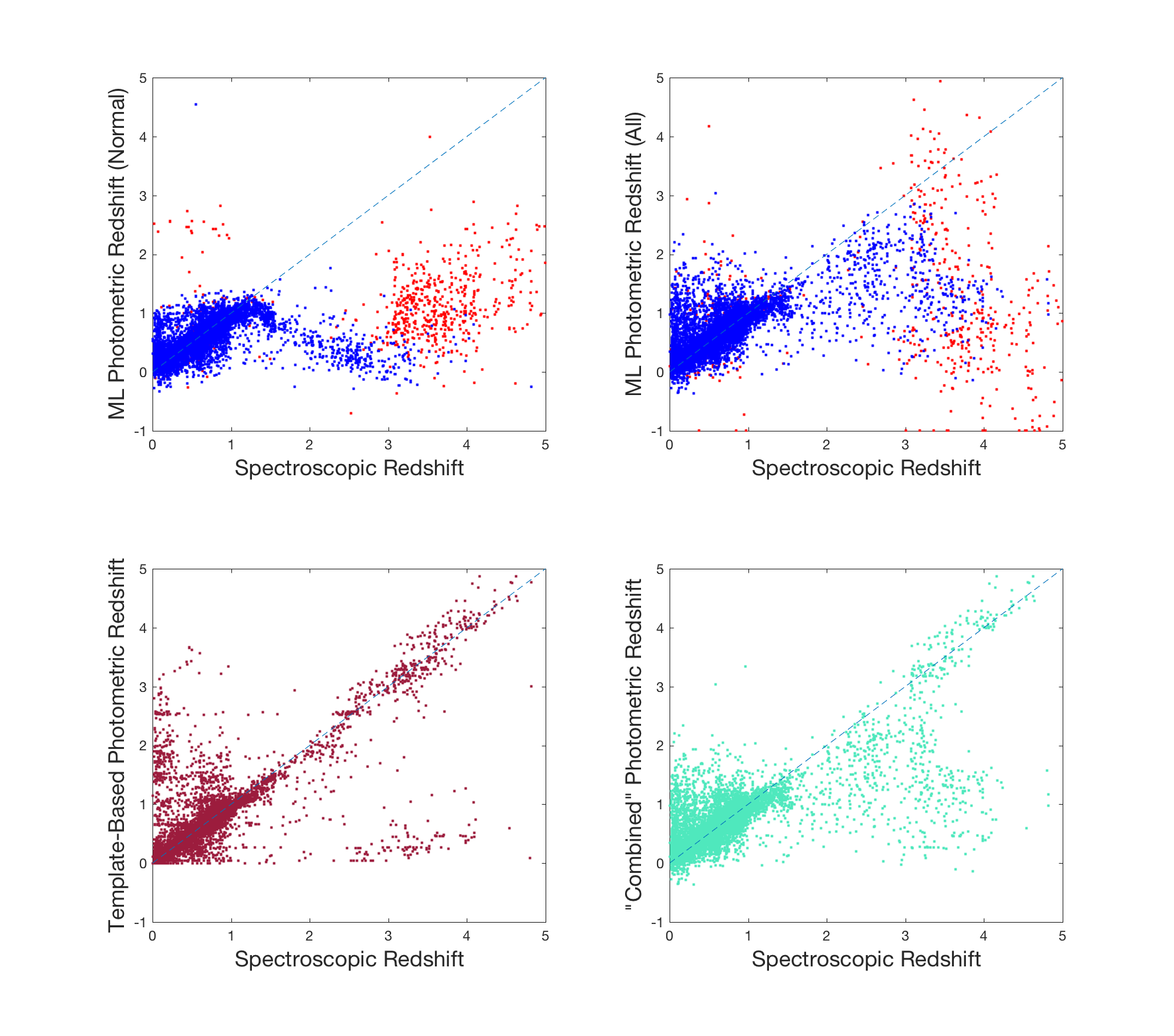}
\caption{Comparison of ML and template methods. Top left: Our `Normal' ML predictions. Top right: Our `All' ML predictions.  Bottom left: The template based predictions of \citet{Adams2020}. Bottom right: The `Combined' predictions that incorporate both the ML and the template based methods. For the plots in the top row, blue indicates galaxies in the interpolative regions of colour-magnitude space, and red those in the extrapolative regions.}
\label{fig:template_method}
\end{figure*}

\begin{figure} 
\includegraphics[scale=0.5]{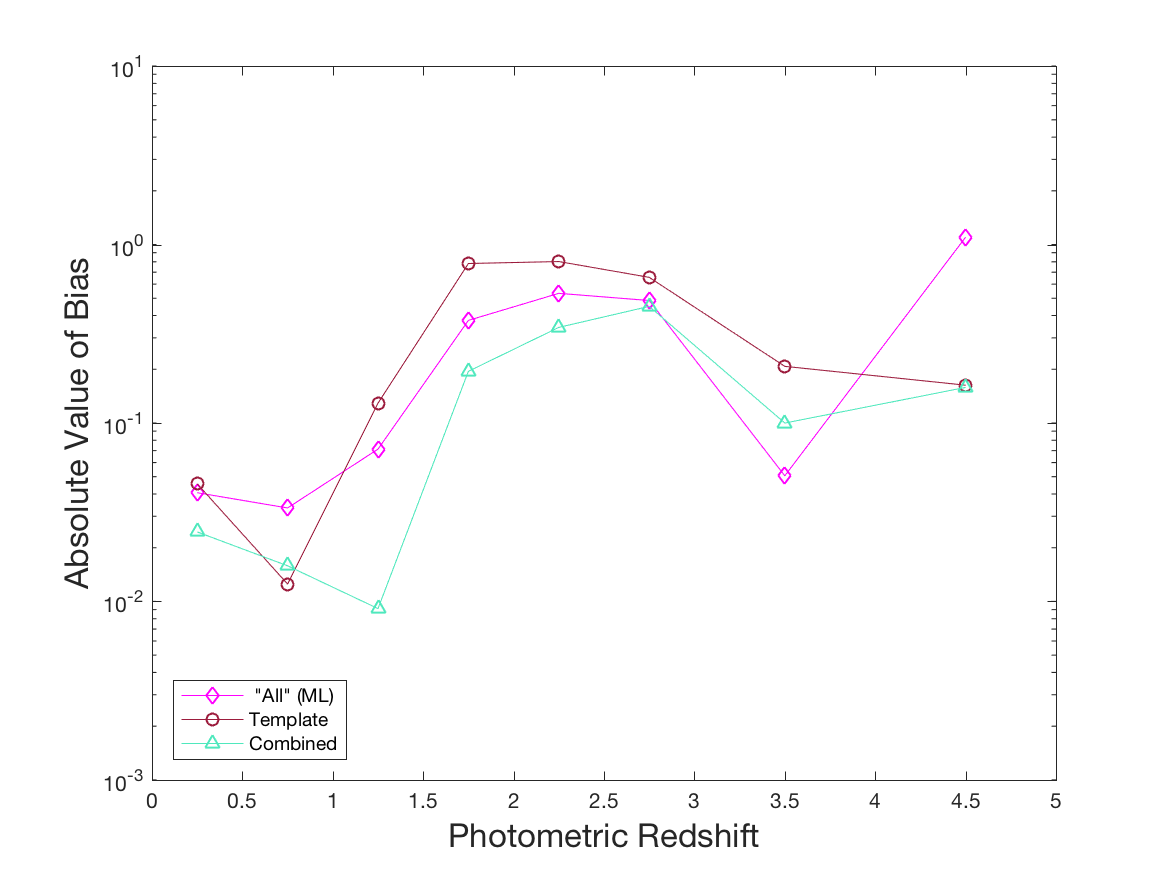}
\caption{Absolute bias as a function of photometric redshift for a) our `All' ML photo-z's, b) the \citet{Adams2020} template-based photo-z's and c) the `Combined' values.}
\label{fig:combination}
\end{figure}

The main focus of this paper is to identify how best to augment machine learning based photo-z's, but it is nonetheless instructive to see how our ML-based predictions compare to template-based methods. In Figure \ref{fig:template_method} we show how our `Normal' and `All' photo-z's compare to the template-based photo-z's calculated in \citet{Adams2020} using {\sc LePhare} (\citealp{Arnouts1999,Ilbert2006}); it can be seen that the ML and the template-based methods each out-perform the other for different galaxies. In particular it can be seen that typically the template-based method is better when GPz is in the extrapolative regime (i.e. where there is little or no training data). The variance in GPz has three components; $\nu$ (uncertainty from lack of data),  $\beta^{-1}_{\star}$ (uncertainty from output noise e.g. galaxies with different redshifts for the same magnitudes) and $\gamma$ (uncertainty from input noise e.g. uncertainty on the magnitudes). The extrapolative uncertainty $\nu$ is typically underestimated (furthermore the transition is smooth with no clear boundary) so we define predictions to be extrapolative if  $\nu>0.1 \times (\beta^{-1}_{\star}+\gamma)$, see \citet{Hatfield2020}. This classification can be used to construct a `Combined' photo-z estimate by using the GPz `All' prediction if GPz is in the interpolative regime, and the {\sc LePhare} photo-z if GPz is in the extrapolative regime.  In Figure \ref{fig:combination} we show the bias of the `Combined' method compared with the ML and template based methods, demonstrating that the `Combined' approach as expected outperforms the individual approaches. For the data sets used here, 4 percent of the test data is in the extrapolative regime, although this fraction typically will vary depending on training and test data used.  

Euclid and Rubin photometry will cover similar wavelengths to this study, over much larger areas (\citealp{Rhodes2017}). Future work will test these methods on the Euclid and Rubin data challenges, combining the pdfs constructed here with template based pdfs using both the interpolative/extrapolative method described here, and the Hierarchical Bayesian Model approach of \citet{Duncan2018}.

\section{Conclusions} \label{sec:conclusions}

In this work we began by constructing mock training and test data sets from CFHT, VISTA, and HSC data, both with spectroscopic redshifts, but with different colour and redshift distributions. This mimics the real issue of spectroscopic training sets having a different colour-magnitude distribution to the target distribution. We then discuss and illustrate several ways of using a combination of GMMs and the machine learning photometric redshift code GPz to obtain improved results over the baseline performance: i) weighting the data appropriately for the colour-magnitude distribution differences, ii) modelling different populations separately, iii) using resampling methods and iv) making the validation data closer to the test data. We compare various metrics from the different methods, finding that respectable improvements in bias at higher redshifts ($z\gtrsim1.5$) can be achieved with these relatively simple methods (and with no additional training data). In particular, modelling different parts of colour-magnitude space separately (`Divide'), Resampling, and weighting the validation data to look more like the test data seem to be the most effective and practical methods. These methods worked well even without removing AGN from the samples.

The key conclusions of this work are:
\begin{itemize}
\item Weighting schemes that take into account the different colour-magnitude distributions of galaxies in the training and test sets can reduce some of the bias in redshift estimation, particularly at high redshift (without any additional data)
\item Using GMMs can help speed up photometric redshift calculation and give improved precision for machine learning based photometric redshift calculation
\end{itemize}

\section*{Acknowledgements}

PH acknowledges funding from the Engineering and Physical Sciences Research Council, generous support from the Hintze Family Charitable Foundation through the Oxford Centre for Astrophysical Surveys, and acknowledges travel support provided by STFC for UK participation in Rubin through grant ST/N002512/1. IAA would like to acknowledge the support of King Abdulaziz City for Science and Technology. ZG is supported by a Rhodes Scholarship granted by the Rhodes Trust. NA acknowledges funding from the   Science and Technology Facilities Council (STFC) Grant Code ST/R505006/1. RAAB acknowledges support from the Glasstone Foundation. This publication arises from research funded by the John Fell Oxford University Press Research Fund.

Based on data products from observations made with ESO Telescopes at the La Silla or Paranal Observatories under ESO programme ID 179.A- 2006. Based on observations obtained with MegaPrime/MegaCam, a joint project of CFHT and CEA/IRFU, at the Canada-France-Hawaii Telescope (CFHT) which is operated by the National Research Council (NRC) of Canada, the Institut National des Science de l'Univers of the Centre National de la Recherche Scientifique (CNRS) of France, and the University of Hawaii. This work is based in part on data products produced at Terapix available at the Canadian Astronomy Data Centre as part of the Canada-France-Hawaii Telescope Legacy Survey, a collaborative project of NRC and CNRS.

The Hyper Suprime-Cam (HSC) collaboration includes the astronomical communities of Japan and Taiwan, and Princeton University. The HSC instrumentation and software were developed by the National Astronomical Observatory of Japan (NAOJ), the Kavli Institute for the Physics and Mathematics of the Universe (Kavli IPMU), the University of Tokyo, the High Energy Accelerator Research Organization (KEK), the Academia Sinica Institute for Astronomy and Astrophysics in Taiwan (ASIAA), and Princeton University. Funding was contributed by the FIRST program from Japanese Cabinet Office, the Ministry of Education, Culture, Sports, Science and Technology (MEXT), the Japan Society for the Promotion of Science (JSPS), Japan Science and Technology Agency (JST), the Toray Science Foundation, NAOJ, Kavli IPMU, KEK, ASIAA, and Princeton University.

This paper makes use of software developed for the Large Synoptic Survey Telescope. We thank the LSST Project for making their code available as free software at \url{http://dm.lsst.org}.

This paper is based, in part, on data collected at the Subaru Telescope and retrieved from the HSC data archive system, which is operated by Subaru Telescope and Astronomy Data Center at National Astronomical Observatory of Japan. Data analysis was in part carried out with the cooperation of Center for Computational Astrophysics, National Astronomical Observatory of Japan.

\section*{Data Availability}

The derived data generated in this research will be shared on reasonable request to the corresponding author.

\bibliographystyle{mn2e_mod}
\bibliography{photoz}

\bsp

\label{lastpage}

\end{document}